\documentclass[12pt,preprint]{aastex}

\def\simgr{\,\hbox{\hbox{$ > $}\kern -0.8em \lower 1.0ex\hbox{$\sim$}}\,}
\def\simle{\,\hbox{\hbox{$ < $}\kern -0.8em \lower 1.0ex\hbox{$\sim$}}\,}
\def\beq{\begin{equation}}
\def\eeq{\end{equation}}

\slugcomment{Revised draft}

\shorttitle{The lithium test in the ONC}
\shortauthors{Palla et al.}

\begin{document}

\title{Age spreads in star forming regions: \\ The lithium test in 
the Orion Nebula Cluster\thanks{Based on data collected at ESO-VLT,
Paranal Observatory, Chile (ID 072.D-0019(B))}}

\author{Francesco Palla and Sofia Randich}
\affil{INAF--Osservatorio Astrofisico di Arcetri, L.go E. Fermi, 5,
I-50125 Firenze, Italy}

\email{palla@arcetri.astro.it,randich@arcetri.astro.it}

\and

\author{Ettore Flaccomio and Roberto Pallavicini}
\affil{INAF-Osservatorio Astronomico di Palermo ``G.S. Vaiana", Piazza del
Parlamento 1, I-90134 Palermo, Italy} 

\email{ettoref@oapa.astropa.unipa.it,pallavic@oapa.astropa.unipa.it}

\begin{abstract}
We present the initial results of a study of the surface lithium abundance in
a sample of low-mass members (M$_\ast \sim$~0.4--1.0 M$_\odot$) of the Orion
Nebula Cluster (ONC) that provide an independent clock to estimate stellar 
ages. We report discovery of significant depletion of lithium in four stars
with estimated mass of $\sim$0.4 M$_\odot$ and age $\sim$10~Myr.  Comparison
with the predictions of numerical and analytical models shows excellent
agreement between the isochronal age and lithium depletion time scale for two
objects, the first case in lithium-poor pre-main sequence stars. Our results
bear on the issue of the real age spread in the ONC and hence on the overall
duration of the star formation process, indicating that the stellar
population did not came into existence in a single, rapid burst.
\end{abstract}


\section{Introduction}

Stars form from the gravitational collapse of dense cores within molecular
cloud complexes.  Whether such clouds can sustain the production of
stars for an extended period of time ($t_{\rm cl}\approx 10^7$~yr), longer
than the typical free fall time ($t_{\rm ff}\approx 10^6$~yr) is still
unknown and critically debated (e.g. MacLow \& Klessen 2004, Tassis \&
Mouschovias 2004).
Empirical information on the duration comes from studies of
the evolution of the gas in molecular cores (the stellar progenitors) and of
the stellar population associated with young clusters and associations.  
%In the former case, the presence of residual atomic gas (HI) well mixed with 
%the dominant molecular gas in several dark clouds indicates that the basic
%process of HI-to-H$_2$ conversion takes place on time scales 
%($\sim$3-10 Myr) longer than previously estimated (Goldsmith \& Li 2005).
As to young stars, the reconstruction of the history of stellar births can be
obtained from the isochronal ages using the HR diagram. Application to
several nearby clusters and associations has shown a common behavior in which
stars begin to form at modest levels roughly 10 Myr in the past, and then at
an accelerating rate with typical e-folding times of 1--3 Myr.  In the case
of the ONC, detailed studies of the optically visible population have
revealed that the period of most active formation is
confined to a few Myr, and has recently ended with the dispersal of the
remnant molecular gas from the winds and UV radiation of the massive
Trapezium stars (Hillenbrand 1997, H97; Palla \& Stahler 1999, PS99; M\"uench
et al.  2002; Flaccomio et al. 2003).

In addition to providing an average age of the ONC, the pattern of stellar
births has revealed another important aspect regarding its {\it age spread}:
the existence of a small, but statistically significant population of
older stars with estimated isochronal ages in excess of $\sim$5~Myr (see also
Slesnick et al. 2004).  Since the existence of older stars in young clusters
bears directly on the issue of the duration of star formation, the
identification of such a population requires  careful examination. 
%especially
%in light of the uncertainties associated with observational parameters as
%well as evolutionary tracks and isochrones (Hartmann 2001). 
In this Letter,
we present the initial results of a study aimed to determine the age spread
of the ONC using measurements of the lithium (Li)
abundance in stars of a selected mass range by means of the Li~I~670.8 nm
doublet.

%\section{Lithium depletion in young stars}

The Li test rests on the ability of stars to deplete their initial
Li content during the early phases of pre--main-sequence (PMS)
contraction. 
%Young, low-mass mass stars ($M _\ast
%\lesssim$1~M$_\odot$) begin their PMS phase with the full interstellar supply
%of Li since the central regions are too cold to ignite nuclear burning
%during the protostellar phase.  As contraction proceeds, the critical
%temperature for Li reactions ($\sim 3\times 10^6$~K) is reached, and the
%initial content is readily depleted in fully convective, sub-solar stars
%(M$\lesssim$0.5 M$_\odot$). Objects with mass $\lesssim$0.065 M$_\odot$ never
%attain the ignition temperature and therefore maintain their initial
%abundance (D'Antona \& Mazzitelli 1994).
It has been shown that the physics required to study the depletion history as
a function of age has little uncertainty, since it depends mostly on 
effective temperature ($T_{\rm eff}$) for fully convective objects (Bildsten 
et al.
1997).  Detailed numerical models show that stars in the
range 0.5--0.2 M$_\odot$ start to deplete Li after about 5--10~Myr, and
completely destroy it after addtional $\sim$10~Myr (e.g., Baraffe et al.  1998, Siess et
al. 2000).  
%Lower mass stars take much longer to burn Li and, due to the
%strong temperature sensitivity of the energy generation rate, there is a
%sharp transition between fully depleted objects and those with the initial
%Li content (the so-called lithium depletion boundary, LDB).  The mass,
%or luminosity, at which the boundary occurs in a cluster reveals its age(e.g.
%Basri 2000).
%This method has been successfully applied to a number of relatively young
%open clusters, such as $\alpha$~Per, IC~2391, NGC~2547, and the Pleiades
%($t\sim 30-120$~Myr), with significant revisions of their ages (e.g.,
%Stauffer et al. 1999, Jeffries \& Oliveira 2005).  
The first evidence for complete Li-depletion 
in a pre--main-sequence (PMS) star has been
reported by Song et al. (2002) in the case of the HIP~112312 binary system.
Similarly, the spectra of both components 
of the spectroscopic binary system St~34, a candidate member of the 
Taurus-Auriga association, 
show no Li absorption line (White \& Hillenbrand 2005). In
both instances, the Li depletion timescale inferred from stellar models
results larger than the isochronal age. 
%thus confirming the discrepancy
%between the two age dating methods found in the above open clusters.
Significant Li depletion was also found in several T Tauri stars at levels
inconsistent with their relatively high luminosities, i.e. young ages
(Magazz\'u et al. 1992, Mart\'\i n et al. 1994).

In spite of these sparse results, the Li test has never been tried
systematically in young clusters in the assumption that subsolar members
would have not had time to deplete significantly their initial
Li content. However, in the case of the ONC the presence of a number of
older stars in the HR diagram suggests that the opposite might be true.  To
verify this, we have selected a sample of 84 low-mass stars in the range
$\sim$0.4--1.0~M$_\odot$ and isochronal ages greater than $\sim$1~Myr (PS99),
drawn from the H97 survey with membership probability
greater than 90\%, based on various proper motion studies. Their distribution
in the HR diagram is shown in Figure~1, together with the region of partial
(light shading) and full (dark shading) depletion predicted by stellar
evolution models (Siess et al. 2000).

\section{Observations and data reduction}

The observations were carried out on 15 February 2004 as part of the
Ital-FLAMES GTO observing time ``Multiobject Spectroscopy of Galactic Open
Clusters of Different Ages and Metallicities" using FLAMES mounted on
VLT-Kueyen (UT2). The Giraffe spectrograph and Medusa fiber system were used
in conjunction with the 316 lines/mm grating and order
sorting filter 15 yielding a nominal resolving power R=19,300 and a spectral
coverage from 660.7 to 696.5~nm that includes the Li~I~670.8~nm line.
We observed the same fiber configuration centered at RA(2000)=05h~35m~16.0m
and DEC(2000)=$-$05d~24m~20s in three separate 1 hr exposures.
%We allocated 84 and 15 fibers on the target stars and on blank sky positions,
%respectively.  
Typical S/N ratios range between $\sim 230$ and $35$ per pixel
for the brightest (I$\sim$13.5) and faintest (I$\sim$17) objects. 
%Flat field, bias, and Thorium-Argon lamp observations were obtained
%during the day before the start of the observing night.

Data reduction was done using the Giraffe BLDRS pipeline\footnote{version 
1.08 -- http://girbldrs.sourceforge.net/}, following the
standard procedure and steps (Blecha \& Simond 2004).  Sky subtraction was
performed separately using the method employed by Jeffries \& Oliveira
(2005).  
%Namely, we created a master sky image by averaging three images
%obtained as the median of groups of five sky spectra.  As in the case of
%Jeffries \& Oliveira, 
We obtained a good sky continuum subtraction, while the
subtraction of the very strong emission lines due to the nebular emission in
the ONC was rather poor.  However, this does not affect the measurement of
the equivalent widths (EW) 
%of the Li line and of the lines used for
%veiling determination (see below) 
since all the lines used in this study are far enough in
wavelength from sky emission.  A sample of spectra in the vicinity of 671~nm
are displayed in Figure~2 for five stars of the same spectral type (M1), but
different luminosity. 
%From these spectra, one can clearly see 
Notice the large
variation of the Li~670.8~nm line from star to star that, in some
case, corresponds to a Li abundance lower than the interstellar value,
as we now show.

%As first steps of the reduction process
%master bias and flat field calibration files are
%produced,  a table containing the position on the CCD
%of the fiber spectra is created from the master flat field together
%with one-dimensional flat field spectra, and a full wavelength solution
%is determined using the ThAr arc spectrum. Then object spectra
%are extracted and subsequently rebinned. The extraction process
%includes bias and scattered light subtraction, localization
%adjustment, actual slit extraction and narrow flat fielding on 
%extracted spectra (for additional details see 
%Blecha \& Simond (2004).

%
\section{Derived lithium abundances}
Our spectra are severely affected by spectral veiling, which needs to be
corrected before determining Li-abundances from the measured EW(Li). Given
$r$, the ratio of the excess to the photospheric continuum, the relationship
between the true and measured EW is:  EW$_{\rm true}$=EW$_{\rm
meas}\;(1+r)$.  In order to estimate $r$, we 
have measured the EW of three strong
lines included in our spectral range 
(Ni~{\sc i}~664.3~nm, Fe~{\sc i} 666.3~nm, and V~{\sc i}~662.5~nm)
in all targets stars and compared
them with those measured in the spectra of stars of similar temperature of
IC~2391 and IC~2602 that are old enough (30--50 Myr, Randich et al. 2001) 
to ensure that their spectra are not affected by veiling. 
%We have 
%selected the two lines used by Magazz\`u et al. (Ni~{\sc i}~664.3~nm and
%Fe~{\sc i} 666.3~nm), plus V~{\sc i}~662.5~nm.  
For each line and each star
in ONC, the quantity $r$ was calculated from EW$_{\rm IC}$/EW$_{\rm ONC}-1$,
where EW$_{\rm IC}$ refers to the open clusters. Finally, the average
value of the veiling from the three lines was computed.  In a few cases it
was not possible to measure the EW of the three lines, and the veiling was
derived using one or two lines only.  
%As expected for young stars, we found
%rather high veiling values ($r\sim 0-8$), with a median $r_{\rm med}$=0.5.
%This is consistent with the results obtained by Hartigan et al. (1995) for a
%large group of T Tauri stars that show both high values (up to $r\sim 9$),
%significant dispersion, and a correlation with the NIR excess.

%Equivalent widths of the Li~670.8~nm line were measured using MIDAS by direct
%integration with respect to the continuum (or pseudo-continuum for the
%coolest objects in our sample). 
Measured Li EWs were then corrected for
veiling using the $r$-values determined as above.  Corrected (true)
EWs are comprised in the range 400--850~m\AA.  Finally, lithium
abundances were derived using MOOG (Sneden 1973) and Kurucz (1993) model
atmospheres, and assuming effective temperatures from the spectral types of
H97 following Hillenbrand \& White (2004). The resulting values of
$\log$~n(Li) ($\equiv$N(Li)/N(H)+12) for the whole sample vary between 2.3
and 3.7, without considering upper limits, with a median value n(Li)=3.12,
very close to the lower limit of the initial interstellar value. 
Random errors in Li abundances include uncertainties in the
correction for veiling, in the measurement of the EW, and in
T$_{\rm eff}$.  We estimate that in most cases the total error in
$\log$~n(Li) does not exceed 0.3~dex.
In order to
estimate errors in the abundance scale, in particular for the coolest stars,
we have also determined abundances using the curves of growth of 
Zapatero Osorio et al. (2002) which yield abundances
systematically lower by $\sim$0.15~dex.

Considering the subsample of stars with mass in the interval $\sim$0.6--1.0
M$_\odot$ (see Fig.~1), in all cases we do not find indication of
Li-depletion.  Actually, the median abundance (n(Li)=3.28) is slightly higher
than that of the whole sample. 
%This result is similar to that found in the T Tauri stars of
%Taurus-Auriga (Magazz\'u et al. 1992, Mart\'\i n et al. 1994). 
The fact that the four stars with isochronal ages greater than 10 Myr and
mass between 0.7 and 0.8 M$_\odot$ display no Li depletion is not too
surprising given the uncertainties introduced by the rapidly developing
radiative core that make the analysis of Li depletion very sensitive to the
detailed physics adopted (convective/radiative boundary and mixing; e.g.
Piau \& Turck-Chi\`eze 2002).

The situation improves for the subsample of stars with estimated mass
below $\sim$0.6 M$_\odot$.  The derived Li abundances are displayed in Figure 3
where it can be seen that the younger ONC stars have maintained all the
initial Li. The exception is the star \#73 for which we could
only derive a lower limit since the poor S/N did not allow a determination of
veiling.  In the case of stars with isochronal ages $\simgr$3 Myr, four of
them show significantly depleted Li-abundances (up to a factor of 7 for star
\#3087), while the rest (8 objects) fall within the region of no depletion.
Table~1 summarizes the properties of the depleted stars. We stress that these
objects have 99\% membership probability based on proper motion (H97), and
the radial velocities derived by us (listed in Table~1) are
consistent with that of the ONC (V$_{\rm r}\simeq$~26.5 km~s$^{-1}$, e.g.
Sicilia-Aguilar et al.  2005).  Given the significance of this finding, in
the following we concentrate the analysis on this restricted group of stars.

\section{Comparison with Models and Implications}

The existence of several Li-poor, low-mass stars in the ONC allows a
direct comparison with both theoretical and analytical models of early
nuclear burning. 
%in an important mass range that has not been tested in
%previous studies of the LDB. 
The mass and isochronal age for each star in Table~1
(labeled HRD) are determined using the PMS models of PS99.  Errors on both
quantities have been estimated from the uncertainties on 
luminosity ($\pm$0.15 dex) and $T_{\rm eff}$ ($\sim \pm$80~K,
corresponding to half a spectral class). The same values (to within
few percent) for the mass are obtained using the PMS models of Siess et al.
(2000), while the ages differ by less than 10\% (larger values for Siess et
al.). Similar agreement is also found using D'Antona \& Mazzitelli (1998)
models, computed with a different treatment of convection.  On the other
hand, the models of Baraffe et al. (1998) yield significantly higher
mass, age and depletion levels.

It is instructive to compare these results with the analytic estimates
derived by Bildsten et al. (1997) that apply to the mass range considered
here. For fully convective objects undergoing gravitational contraction at
approximately constant effective temperature and assuming fast and complete
mixing, Bildsten et al. derive simple and accurate (better than $\sim$20\%)
relations for the time variation of the luminosity (their eq. 4) and for the
amount of Li depletion at a given age (their eq. 11).  From the observed
$L$ and $T_{\rm eff}$, one can then construct a plot of the mass-depletion
time relation that is bounded by the determination of the Li abundance.
The results for the four depleted stars are shown in Figure~4. In the
diagram, the observed luminosity is represented by a line with positive
slope, while the depletion time is set by the observed Li abundance
represented by a line with negative slope. The intersection of the two lines
yields the combination of mass and age at the given $T_{\rm eff}$.  The last
two columns of Table~1 list the derived values obtained in this manner. The
hatched region in each panel represents the allowed region bound by the
uncertainties in $L$ and in n(Li) ($\pm$0.3 dex).

Stars \#3087 and 775 have almost identical values of $T_{\rm eff}$ and $L$,
and therefore occupy the same position in the plot. In both cases, we find 
excellent agreement between the observed and theoretical estimates: the
stellar mass is well represented by the evolutionary tracks (to within 10\%),
while the isochronal age differs from the depletion age by 10\%. To our
knowledge, this is the first time that the two methods give consistent
results to such high accuracy. The agreement can be improved by considering
that the age at a given depletion is quite sensitive to $T_{\rm eff}$
($t\propto 1/T_{\rm eff}^{3.4}$). Thus, a small variation of this quantity
shifts the position of the hatched region.  For example, using a 
hotter temperature scale (by $\sim$50 K), the resulting
difference between the isochronal and depletion ages would be reduced to less
than 5\%, while the mass would maintain the same consistency. Thus, our
observations not only confirm that Li can serve as a powerful and highly
accurate age clock in young clusters, but also as a test of PMS models.

The situation for the two other stars displayed in Figure~4 is not as
favorable as in the previous case. In both stars \#674 and 335, the derived
n(Li) yield ages that are marginally inconsistent with the isochronal ones
(at the 1-$\sigma$~level): these stars
have experienced too little burning for the estimated ages. 
However, both
objects are within the bounded region predicted by Bildsten et al. analysis.  
%Contrary to the case of HIP~112312A and St~34 mentioned in
%Sect.~2, here we do find
%that the isochronal ages are systematically too old. 
A combination of cooler
$T_{\rm eff}$ and/or higher $L$ would help in reducing the discrepancy, thus
allowing the stars to fall within (or very close to) the hatched regions.

We have already noticed that other stars that should have shown some level of
Li depletion actually do not pass the test (see Fig. 3). In particular,
star \#114 has n(Li)=3.18 although its luminosity
and effective temperature are almost identical to those of \#775 and 3087.
Less problematic is the case of star \#625 (n(Li)=3.17)
with the same $T_{\rm eff}$ but higher luminosity (log $L=-$0.73) that places
it right at the boundary of the depletion region (cf.  Fig.~1). Of the two
stars with log $T_{\rm eff}$=3.568 (spectral type M0.5), \#277 shows only
marginal Li depletion (n(Li)=2.96) despite an isochronal age of $\sim$10
Myr. Finally, the only object with mass $\sim$0.57 M$_\odot$ that falls
within the theoretical region of full depletion actually displays the initial
amount of Li.  However, at this mass and age such an object has already
developed a small radiative core, so again the negative result is not
completely unexpected.

\section{Conclusions}

The initial study of the Li abundance in the ONC has uncovered four low-mass,
{\it bona-fide} members that show partial depletion at levels consistent with
their age and mass.  The derived ages of about 10 Myr indicate that the ONC
does contain objects much older than the average age of the dominant
population.  The identification of such a group of stars has significant
implications for the star formation history of the cluster, indicating that
its duration extends long in the past, although at a reduced rate, in
accordance with the empirical evidence found in the majority of nearby star
forming regions (Palla \& Stahler 2000). We note that the presence of a
population in the ONC with isochronal ages in excess of $\sim$10 Myr has been
independently inferred by Slesnick et al. (2004) in their study of the very
low-mass and brown dwarf members. However, this apparently old population
could also represent a population of scattered light objects due to disk
occultation whose true luminosity has been understimated. In our case, the
consistency between the Li and isochronal age for the Li-depleted stars makes
this intepretation unnecessary.

Finally, we like to emphasize that it is important to extend the present
observations to lower mass members of the ONC where the transition between
depleted and undepleted stars should occur.  The study of H97 allows us to
probe the critical mass interval between $\sim$0.4 and $\sim$0.2 M$_\odot$,
while lower mass objects would be too faint for high spectral resolution
observations in the optical. However, considering that a 0.3~M$_\odot$ star
is expected to deplete half of its initial Li in more than 15 Myr, it is
possible that the observations presented here might have uncovered the most
Li depleted stars of the ONC.

{}
\clearpage

\begin{deluxetable}{rcccccccc}
\tablecaption{Properties of the Li-depleted stars\label{tbl-1}}
\tablewidth{0pt}
\tablehead{
\colhead{ID} & \colhead{log L} & \colhead{log T$_{\rm eff}$} 
& \colhead{V$_{\rm r}$} 
& \colhead{n(Li)} & \colhead{M$_{\rm HRD}$} & \colhead{t$_{\rm HRD}$} 
& \colhead{M$_{\rm Li}$} & \colhead{t$_{\rm Li}$}\\
 & (L$_\odot$) & (K) & (km/s) &  & (M$_\odot$) & (Myr) & (M$_\odot$) & (Myr) 
}
\startdata
775  & $-$1.006 & 3.557 & 31.6 &2.44 & 0.39$\pm$0.07 & 9.7$\pm$3.9  & 
0.44$\pm$0.06 & 10.8$\pm$1.5 \\
3087 & $-$1.014 & 3.557 & 27.5 &2.27 & 0.39$\pm$0.07 & 9.8$\pm$3.8  & 
0.44$\pm$0.06 & 11.0$\pm$1.5 \\
335  & $-$1.206 & 3.557 & 26.5 &2.67 & 0.39$\pm$0.07 & 18.4$\pm$4.0 & 
0.33$\pm$0.06 & 12.5$^{+2.0}_{-1.4}$ \\
674  & $-$0.999 & 3.568 & 26.6 &2.88 & 0.46$\pm$0.08 & 15.6$\pm$5.0 & 
0.40$\pm$0.07 & 9.3$^{+2.5}_{-1.4}$ \\
%4 & 277  & $-$0.919 & 3.568 & 3.06 & 0.48$\pm$0.08 & 10.0$\pm$3.9 & 
%0.42$\pm$0.09 & 7.5$^{+2.5}_{-0.8}$ \\
%1 & 53   & $-$0.707 & 3.580 & 3.47 & 0.57 & 6.9  &      &     \\
%5 & 114  & $-$1.013 & 3.557 & 3.29 & 0.39 & 9.7  &      &     \\
%7 & 625  & $-$0.730 & 3.557 & 3.17 & 0.40 & 3.7  &      &     \\
%  & 690  & $-$0.370 & 3.580 & 2.94 & 0.56 & 2.7  &      &     \\
%  & 661  & $-$0.390 & 3.580 & 2.99 & 0.55 & 2.8  &      &     \\
%  & 520  & $-$0.370 & 3.580 & 3.07 & 0.56 & 2.7  &      &     \\
%  & 313  & $-$0.500 & 3.580 & 3.25 & 0.56 & 3.7  &      &     \\
%  &  53  & $-$0.707 & 3.580 & 3.47 & 0.57 & 6.9  &      &     \\
 \enddata

\end{deluxetable}
\clearpage

 \begin{figure}
% \epsscale{.80}
 \plotone{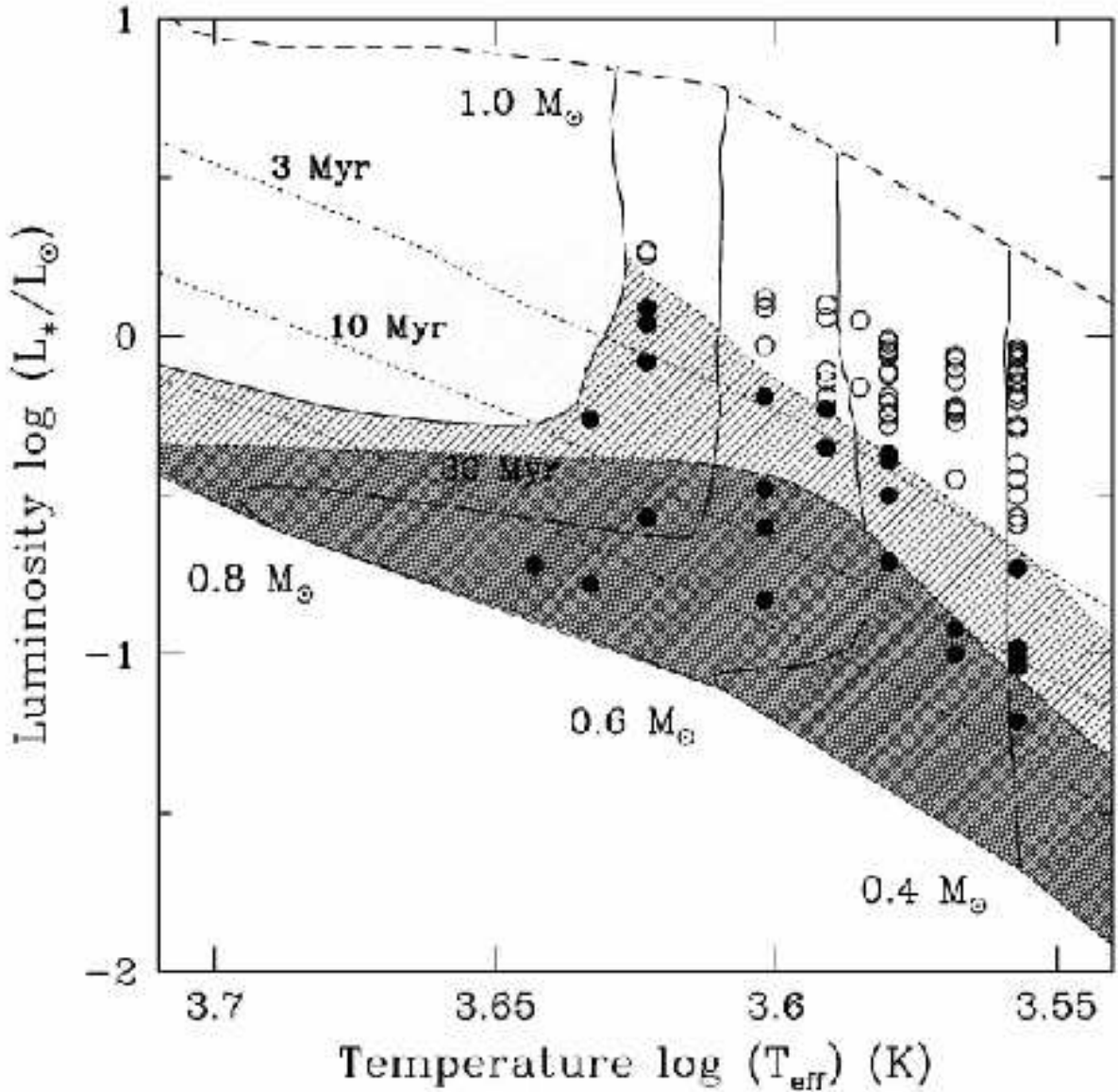}
 \caption{The distribution  of the sample stars in the H-R diagram.
 The hatched
 regions indicate different levels of predicted Li depletion: up to
 a factor of ten (light grey) and more (dark grey) below the initial value
 according to the models of Siess et al. (2000).
 Selected masses and isochrones are indicated. Open and filled circles are
 for theoretically expected undepleted and depleted stars, respectively.
 \label{fig_hrd}}
 \end{figure}

 \begin{figure}
% \epsscale{.50}
 \includegraphics[angle=-90,scale=.85]{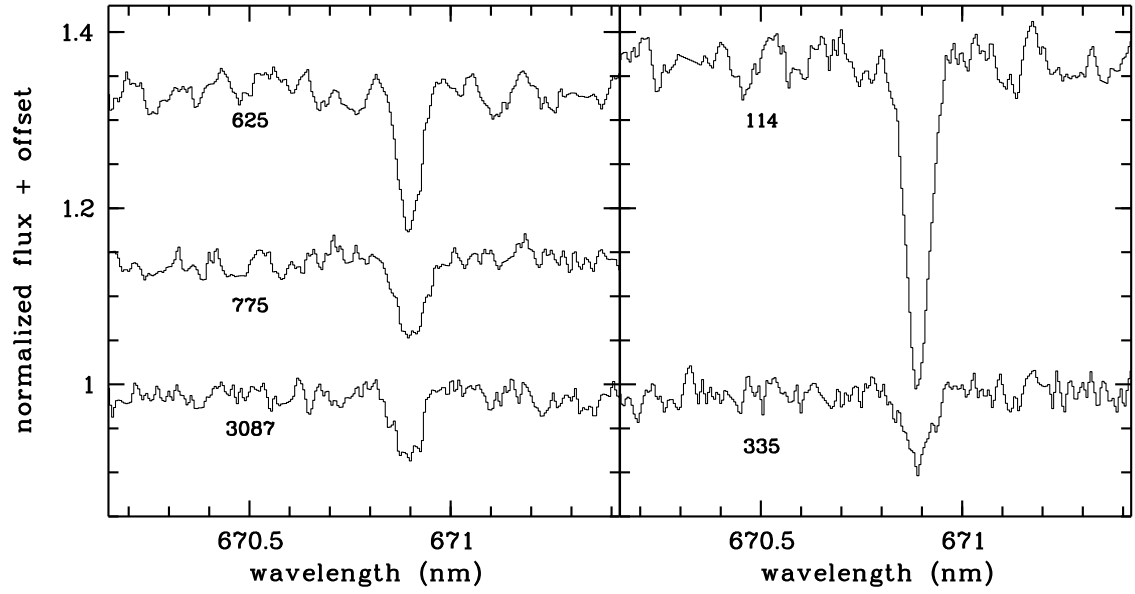}
 \caption{Portion of the VLT-Giraffe spectra in the vicinity of the Li~670.8~nm
 line of the five faintest and coolest stars of the ONC sample (see Fig.~1).
 Numbers refer to the optical star designation in H97. 
%The narrow emission spikes in the spectra \#114, 335, 625 are due to 
%cosmic rays.
 \label{lid_spectra}}
 \end{figure}

 \begin{figure}
% \epsscale{.80}
 \plotone{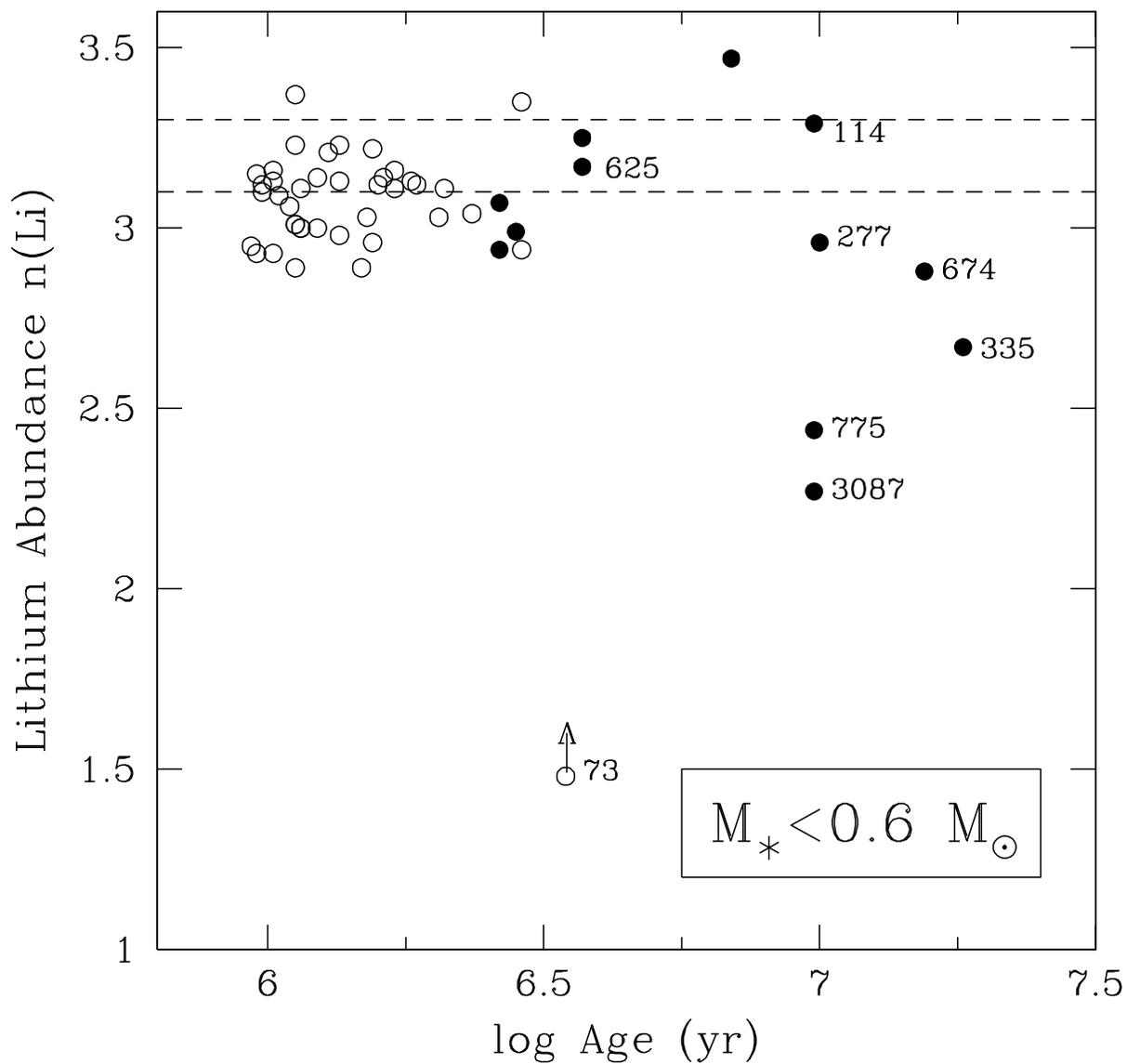}
 \caption{Li abundance vs. age for stars with mass $<$0.6 M$_\odot$. 
 Symbols are the same as in Fig.~1. The dashed horizontal lines mark
 the region of the interstellar Li-abundance (3.1--3.3). 
 Typical errors in n(Li) are of $\pm$0.3~dex. All labeled stars are
 discussed in the text.
 \label{fig_nli}}
 \end{figure}

 \begin{figure}
% \epsscale{.80}
 \plotone{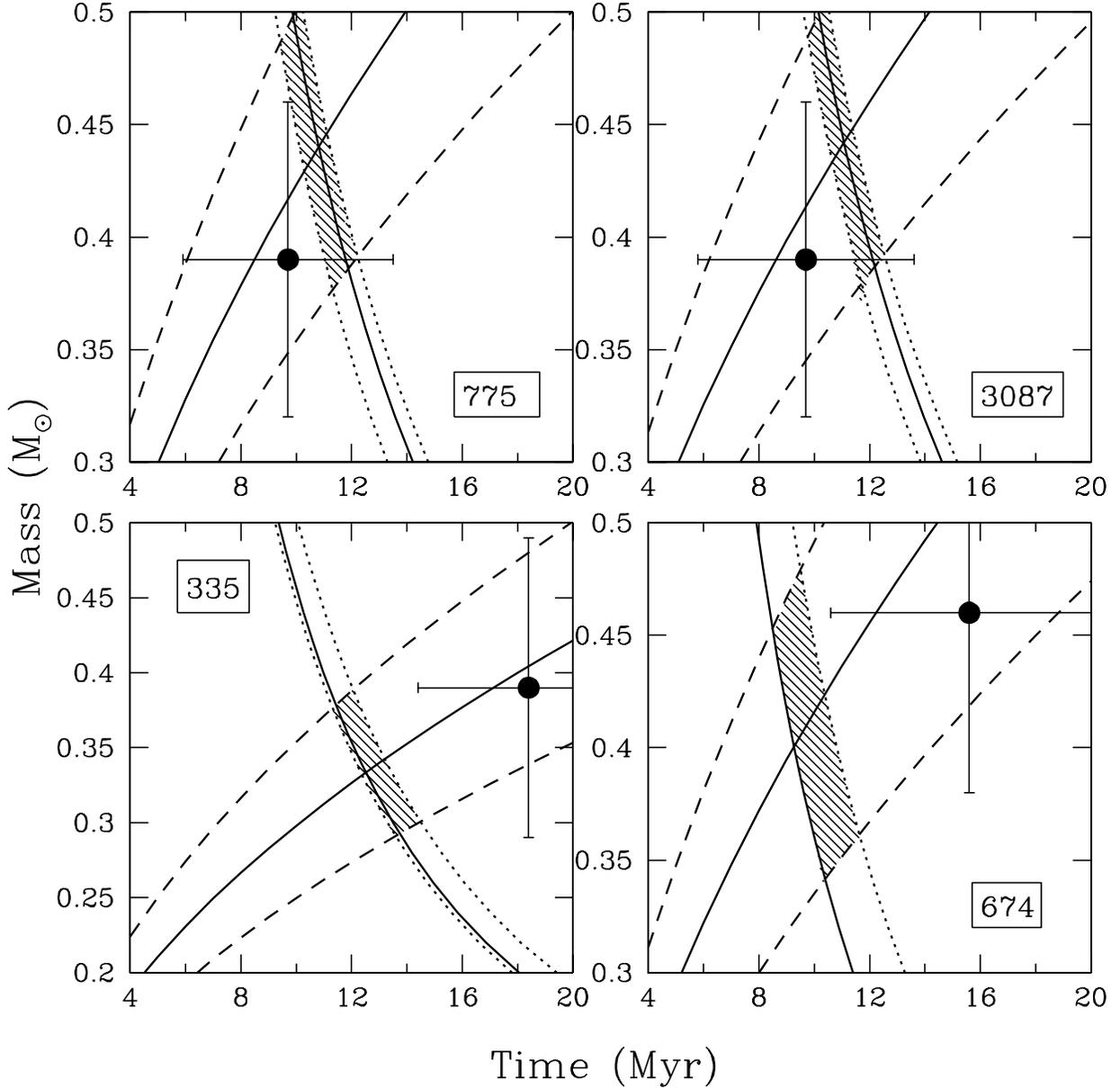}
 \caption{Mass vs. age plot for the four stars with evidence for Li
 depletion. For each star, we plot the luminosity curve (positive slope) and
 the Li abundance curve (negative slope) computed for the value of
 $T_{\rm eff}$ given in Table~1. The dashed curves give the uncertainty range
 in the observed $L$ ($\pm$0.15 dex, long-dash) and in the measured n(Li)
 ($\pm$0.3 dex, short-dash). The hatched region bounds the values of $M$ and
 $t$ consistent with the observations. In each panel, the solid points with
 errorbars give the mass and age from theoretical PMS tracks and isochrones
 (PS99).
 \label{fig_4panels}}
 \end{figure}

\clearpage

\end{document}